\documentclass{optica-article}

\journal{opticajournal} 

\articletype{Research Article}
\usepackage{lineno}
\usepackage{float}

\begin{document}

\title{Ultra-broadband spectral and polarisation entanglement using dispersion-engineered nanophotonic waveguides}

\author{Mahmoud Almassri \authormark{*} and Mohammed F. Saleh}

\address{MNO Group, Institute of Photonics and Quantum Sciences, Heriot-Watt University, EH14 4AS Edinburgh, UK}

\email{\authormark{*} mha2003@hw.ac.uk} 
\begin{abstract}
In this paper, we propose exploiting dispersion-engineered nanophotonic waveguides in generating unprecedented  ultra-broadband spectral and polarisation entanglement using spontaneous four-wave mixing parametric processes. We developed a simplified theoretical quantum framework to investigate and analyse these interactions under pulse source excitations. Using aluminum gallium arsenide and thin-film lithium niobate waveguides, we anticipate to obtain photon pairs with high-dimensional frequency entanglement, characterised by Schmidt numbers as large as ~10$^8$, and covering the wavelength range 940--2730 nm (equivalent to a 210 THz bandwidth). Additionally, we show that Al$_{0.3}$Ga$_{0.7}$As waveguides with hybrid cladding can enable the generation of polarisation-entangled photon pairs with concurrence exceeding 0.93 across wavelengths from 1175 nm to 1750 nm, spanning almost across all the telecommunication bands with only approximately 27 nm window inevitable-degradation around the pump wavelength. We envisage that these introduced integrated on-chip sources will significantly advance quantum photonic technologies, enabling breakthroughs in multi-channel quantum networking and scalable quantum information systems.
\end{abstract}

\section{Introduction}
 Entangled photon pairs are essential for applying quantum theory in communication, computation, and sensing, leading to breakthroughs such as quantum teleportation, quantum secure key distribution, and quantum-enhanced metrology \cite{RevModPhys.84.777,10.1063/1.5115814,wang_integrated_2020,pirandola_advances_2018,reimer_generation_2016,cozzolino_high-dimensional_2019,humphreys_deterministic_2018,degen_quantum_2017}. Among the various degrees of freedom in which entanglement can be obtained, spectral and polarisation entanglement have played crucial roles in enabling quantum technologies. Spectral entanglement has allowed the encoding of high-dimensional quantum states, enhancing information capacity and resilience against decoherence \cite{Law00,cozzolino_high-dimensional_2019}. Meanwhile, polarisation entanglement has been widely implemented in quantum networking applications \cite{Bouw97,OBrien03}.

Periodically-poled bulk crystals and dispersion-engineered optical fibres have been exploited in generating spectral and polarisation entangled photons via accommodating nonlinear spontaneous parametric processes. Highly dimensional broadband spectral entanglement has been achieved using photonic crystal fibers fibres, superlattice structures, and chirped periodically-poled structures \cite{garay-palmett_ultrabroadband_2008,Zhang07,Tashima24}. Similarly, dispersion-shifted fiber loop and periodically poled KTiOPO$_{4}$ crystals have been employed to produce polarised entangled photons \cite{takesue_generation_2004,fiorentino_source_2005}. However, these bulk and fiber-based sources  face challenges in scalability and integration with on-chip quantum technologies.

Integrated waveguides have emerged as a potential alternative technique that offers stable, compact, and scalable entangled-photons sources \cite{OBrien09,Feng20,Matsuda16}. The transition to nanophotonic waveguides provides particularly strong spatial mode confinement, enabling exceptional dispersion engineering flexibility.  Broadband spectral entanglement over 1000 nm frequency range has been demonstrated in dispersion-engineered integrated silicon-nitride waveguides  \cite{vijay_sin_2023}. Meanwhile, polarisation entangled photon pairs with concurrence >0.9 have been achieved in silicon and AlGaAs waveguides \cite{sharma_silicon_2022,Kultavewuti17}. 

In this work, we propose using dispersion-engineered nanophotonic waveguides to produce photon pairs that are entangled in frequency and polarisation over an exceptionally ultra-broadband spectral range. We apply our approach to thin-film lithium niobate (TFLN) and Al$_{0.3}$Ga$_{0.7}$As platforms. The former is characterised by low propagation loss ($<$1~dB/cm), broad transparency range 0.35--5~$\mu$m,  strong second- and third-order,   nonlinearities, and mature fabrication and integration techniques \cite{luo2023advances,jankowski_ultrabroadband_2020,lauria_mixing_2022}. TFLN waveguides have been widely implemented in classical and quantum applications, such as supercontinuum generation~\cite{lu_octave-spanning_2019} and entangled photon sources~\cite{zhao_high_2020}.  On the other hand, Al$_{0.3}$Ga$_{0.7}$As offers CMOS compatibility, co-integration with detectors and modulators~\cite{baboux_nonlinear_2023,autebert_integrated_2016}, ultra-broadband transmission range 0.62--17~$\mu$m, and slightly higher linear losses ($\sim$3~dB/cm) \cite{mahmudlu_algaas--insulator_2021}. 


We exploit and prioritise spontaneous four-wave mixing (SFWM) process over spontaneous parametric down-conversion (SPDC) in generating entangled photon pairs, since SFWM can enable  extremely broader phase-matching bandwidths. Contrarily to SPDC, SFWM also does not require birefringence or implementing quasi-phase-matching schemes to achieve efficient nonlinear interactions. By precisely tailoring the waveguide dimensions to balance material and waveguide dispersions, we achieve, to the best of our knowledge, the broadest phase-matching bandwidth of the SFWM process. The quality of the output spectral and polarisation entanglement are quantified via calculating the Schmidt number and concurrence, respectively.

The paper is organised as follows: Section 2 covers the theoretical model used in investigating SFWM processes under pulse excitations, and the employed methods for calculating the Schmidt number and concurrence. The generation of ultra-broadband frequency-entangled photons is detailed in Sec. 3, and polarised-entangled photons in Sec. 4, both sections thoroughly discuss waveguide designs and output results. Finally, our conclusions are summarised in Sec 5.

\section{Modeling spontaneous four-wave mixing with pulsed-pump excitations}
In a spontaneous four-wave mixing (SFWM) process, two pump photons $p_{1}$, $p_{2}$ with frequencies $\omega_{p_{1}}$, $\omega_{p_{2}}$  are annihilated, generating two other photons signal $s$ and idler $i$ with frequencies $\omega_{s}$, $\omega_{i}$, such that $\omega_{p_{1}}+\omega_{p_{2}}=\omega_{s}+\omega_i$. Under pulsed-pump excitations, multiple SFWM processes can take place provided that the phase-matching condition is satisfied or nearly met \cite{agrawal2007nonlinear}. To model this interaction, we modify our approach that is developed in Refs. \cite{Saleh2019,greenwood_narrowband_2021,almassri_heralded_2024} and based on the Heisenberg equations of motion. For a certain signal and idler pair $\left(\omega_{s},\omega_i\right)$, the spatial evolution of the signal annihilation $\hat{b}_s$, and idler creation   $\hat{b}_i^\dagger$ operators along the propagation direction $z$ are governed by,
$$\frac{\partial\hat{b}_s}{\partial z} = -j\frac{\Delta k}{2}\hat{b}_s + j\gamma\hat{b}_i^\dagger,$$
$$\frac{\partial\hat{b}_i^\dagger}{\partial z} = j\frac{\Delta k}{2}\hat{b}_i^\dagger - j\gamma\hat{b}_s,$$
where $\gamma$ is an effective third-order nonlinear coupling coefficient that is proportional to the sum of the dominant spectral pump components,  $\Delta k = \kappa_{p_{1}}+\kappa_{p_{2}} - \kappa_s - \kappa_i$ is the average phase mismatch associated with the dominant SFWM processes, $\kappa_{q}$ is the photon $q$ propagation constant that incorporates contributions of self- and cross-phase modulations effects \cite{almassri_heralded_2024}. Linear and nonlinear losses can be safely neglected in this analysis, since only short TFLN and AlGaAs waveguides have been considered  \cite{luo2023advances,mahmudlu_algaas--insulator_2021} and the possible two-photon absorption process is highly unlikely in both platforms due to the low intensity of the generated photon pairs \cite{zhu_integrated_2021,may_second-harmonic_2019}. 



The analytical solution of the above differential equations can be written in the form, $$
\left[\begin{matrix} 
\hat{b}_s(z) \\ 
\hat{b}_i^{\dagger}(z) 
\end{matrix}\right] = 
\mathbf{T}(z)
\left[\begin{matrix} 
\hat{b}_{s}(0) \\ 
\hat{b}_{i}^{\dagger}(0) 
\end{matrix}\right],  
$$
where
$$
\mathbf{T}(z) = 
\left[\begin{matrix} 
\cosh(gz) - \dfrac{j\Delta k}{2g}\sinh(gz) & \dfrac{j\gamma}{g}\sinh(gz) \\ 
\dfrac{-j\gamma}{g}\sinh(gz) & \cosh(gz) + \dfrac{j\Delta k}{2g}\sinh(gz) 
\end{matrix}\right],
$$
is the transfer matrix that relates the field operators at position $z$ to their initial values, and $g = \sqrt{|\gamma|^2 - \dfrac{\Delta k^2}{4}}$.

\subsection{Frequency Entanglement}
The joint spectral amplitude (JSA) that characterises  the spectral correlations of the generated photon pairs is determined by the off-diagonal element $T_{12}\left(\omega_s,\omega_i\right)$ of the transfer-matrix $\mathbf{T}$. Using the Schmidt decomposition, $T_{12}$ can be expressed as $ \sum_n \sqrt{\lambda_n} u_n(\omega_s) v_n(\omega_i)$, where $u_n$ and $v_n$ form orthonormal sets of Schmidt modes, and $\lambda_n$ are the Schmidt coefficients \cite{mosley_conditional_2008}. The degree of spectral entanglement is quantified by the Schmidt number defined as   $K = 1/\sum_n \lambda_n^2$. A spectrally disentangled pure state has $K=1$, while a highly entangled state has $K \gg 1$. Calculating $K$ using the singular-value decomposition (SVD) of the JSA matrix can be computationally intensive for GHz or MHz pump sources, since it would require a very fine spectral grid.  In this case, an approximate analytical equation $K \approx \frac{\sigma^-}{\sqrt{2}\,\sigma^+}$ can be used  with $\sigma^-$ and $\sigma^+$ the effective bandwidth parameters related to the phase-matching and pump spectra, respectively \cite{Zhang07}. 

\subsection{Polarisation Entanglement}
For a two-mode waveguide, the output single photons of a SFWM process can be either in TE or TM mode. Nearly phase-matched nonlinear interactions are possible either when all the interacting photons share the same polarisation state or the two pump photons have opposite polarisations and similarly the signal and idler. Hence, the pure output state of the photon pair can be written as:
\begin{equation}
    |\psi_{\text{out}}\rangle = c_{\text{TE-TE}}|\mathrm{TE}_s,\mathrm{TE}_i\rangle + c_{\text{TM-TM}}|\mathrm{TM}_s,\mathrm{TM}_i\rangle + c_{\text{TE-TM}}|\mathrm{TE}_s,\mathrm{TM}_i\rangle + c_{\text{TM-TE}}|\mathrm{TM}_s,\mathrm{TE}_i\rangle,
\end{equation}
where $c$ is the probability amplitude of a certain state. To generate a symmetric Bell state with a maximal polarisation entanglement in the form,
 \begin{equation}
    |\psi\rangle = \frac{1}{\sqrt{2}} \left( |\mathrm{TE}_s,\mathrm{TE}_i\rangle + |\mathrm{TM}_s,\mathrm{TM}_i \right),
    \label{eqn:Bell}
\end{equation}
the other two nonlinear interactions with opposite polarisations should have negligible contributions to the final state. The degree of the polarisation entanglement can be measured via calculating the concurrence \cite{zhang_direct_2013},
\begin{equation}
    C = 2|c_{\text{TE-TE}}c_{\text{TM-TM}} - c_{\text{TE-TM}}c_{\text{TM-TE}}|, 
    \label{eqn:Concurrence}
\end{equation}
where $C=1$ corresponds to the maximal entanglement case  with $|c_{\text{TE-TE}}| = |c_{\text{TM-TM}}| = 1/\sqrt{2}$, and $c_{\text{TE-TM}}, \, c_{\text{TM-TE}} = 0$. The entanglement quality can be degraded by polarisation mode dispersion (PMD) that creates temporal distinguishability between the different polarisation states. The differential group delay between TE and TM polarisations of the photon $q=s,i$ is,
\begin{equation}
\tau_{q} = \frac{L}{v_g^{\left(\mathrm{TE}_q\right)}} - \frac{L}{v_g^{\left(\mathrm{TM}_q\right)}},
\end{equation}
with $v_g$ the group velocity, and $L$ the waveguide length. Assuming having Gaussian filters for all the photons and an ideal phase matching for the frequency range covered by the filters, the effect of the PMD on the concurrence can be analytically expressed as \cite{sharma_silicon_2022}:
\begin{equation}
C(\tau_s, \tau_i) = \exp\left[-\frac{(\tau_s - \tau_i)^2\sigma^2_s\sigma^2_i + \sigma^2_p(\tau^2_s\sigma^2_s + \tau^2_i\sigma^2_i)}{8(\sigma^2_s + \sigma^2_i + \sigma^2_p)}\right],
\end{equation}
where $\sigma_{s(i)}$ are the bandwidths of the signal (idler) filters and $\sigma_p$ is the pump bandwidth. 


\section{Ultra-broadband frequency entanglement using nano-photonic waveguides}
\begin{figure}
\centering\includegraphics[]{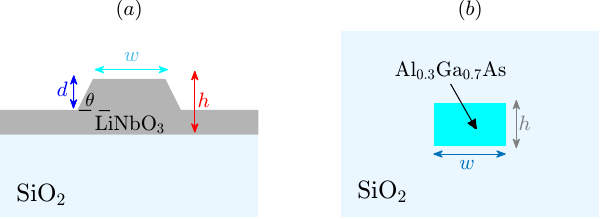}
\caption{(a) Cross-section of a TFLN waveguide on silica, with thickness $h$, top width $w$, etch depth $d$, and side-wall angle $\theta$. (b) Cross-section of an Al$_{0.3}$Ga$_{0.7}$As rectangular waveguide with height $h$,  and width $w$ buried in silica cladding.}
\label{fig:1}
\end{figure}

In this section, we apply the developed model to investigate SFWM in TFLN on silica substrate and Al$_{0.3}$Ga$_{0.7}$As buried in silica nanophotonic platforms, shown in Fig. (\ref{fig:1}). The dispersion profiles of the fundamental modes of the waveguides are engineered to achieve ultra-broadband SFWM phase-matching. Pulsed pumps with various spectral widths (0.5--15 THz) corresponding to a range of temporal widths (4.7--0.16 ps) are considered. The pump central wavelength is approximately at the waveguide zero-dispersion wavelength (ZDW).

\subsection{TFLN waveguides}
\begin{figure}[t]
\centering\includegraphics[width=12.1 cm, height=12.1 cm]{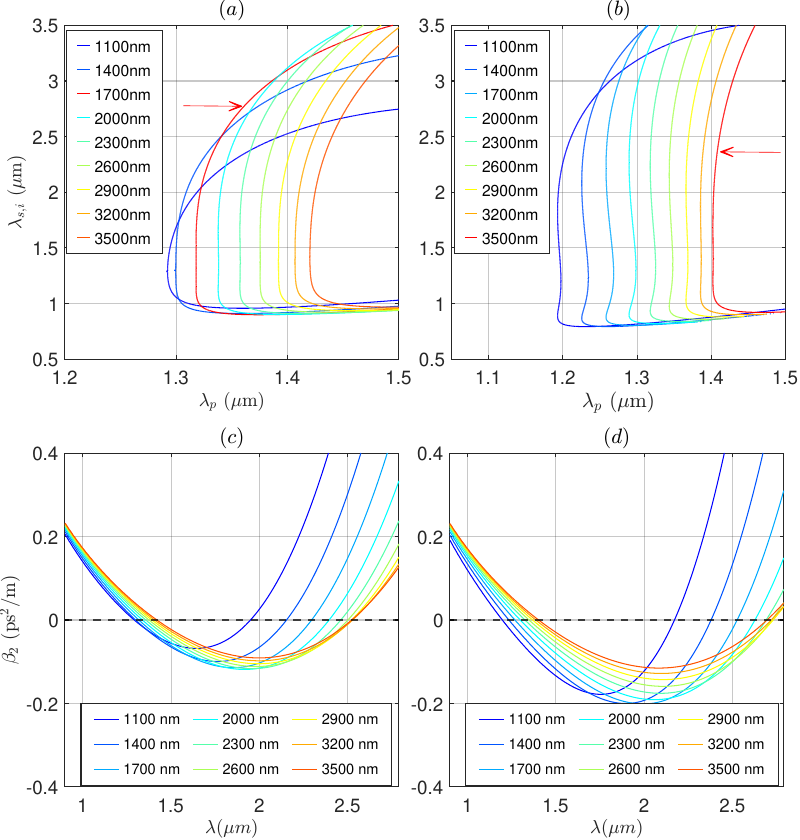}
\caption{ (a,b)  Phase-matching contours of  SFWM processes in  TFLN waveguides with different top widths, and fixed etch depth 300 nm (a) and 433 nm (b). (c,d) Group velocity dispersion $\beta_2$ for the structures employed in (a,b), respectively.
The values shown in the legends are for the different top widths used in the 
simulations.}
\label{fig:2}
\end{figure}

The phase-matching contours of several SFWM processes in TFLN waveguides with different combinations of the etch depth $d$ and top width $w$ are portrayed in Figs. \ref{fig:2}(a,b). The waveguide thickness $h$ and side-wall angle are kept constant at 800 nm, and $\theta=60^\circ$, respectively. All the interacting modes are in the waveguide fundamental $\mathrm{TE}$ modes. Each contour indicates the signal and idler wavelength pairs that satisfy the phase-matching condition $\Delta k = 0$ at a given degenerate pump wavelength. Effective refractive indices of guided modes, incorporating material dispersion, were calculated using COMSOL software.

The phase-matching contour for each waveguide has a characteristic of nearly vertical flat edge segment that indicates a broadband of signal-idler wavelengths can be generated around a particular pump wavelength. For a certain etch depth, as the top width $w$ increases, the bandwidth of the flat segment becomes broader and is shifted to longer pump wavelengths. However, increasing the etch depth $d$ for a given width shifts the contours to shorter pump wavelengths. The associated group-velocity dispersion (GVD) with these flat segments is very small as depicted in Figs. \ref{fig:2}(c,d) that show second-order dispersion coefficient $\beta_2 = d^2k/d\omega^2$. The displayed TFLN waveguides are characterised by 2 ZDWs, with the flat segments displayed in Figs. \ref{fig:2}(a,b) occur at the shorter ZDW.

\begin{figure}[t]
\centering\includegraphics[width=12 cm, height=12 cm]{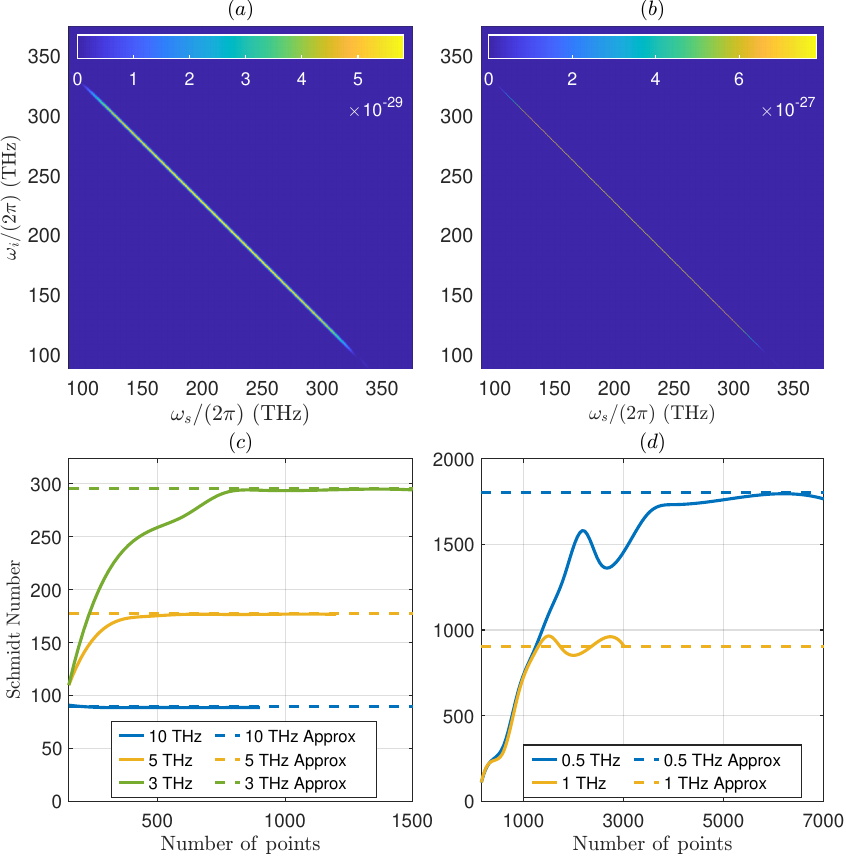}
\caption{(a,b) JSI of SFWM processes in the TFLN waveguide with $433$~nm etch depth  and $3500$~nm top width, using a pump source at 1401.5 nm central wavelength and pump spectral widths  15 THz and 1 THz, respectively. The colorbars indicate the probability density in $\mathrm{Hz^{-2}}$.  (c,d) Dependency of the Schmidt number $K$ on the number of the sampling frequency points that defines the x-axis in panels (a,b) for different input pump bandwidths. The rest of the simulation parameters are the same as in (a,b). }
\label{fig:3}
\end{figure}

The joint spectral intensities (JSI) of SFWM interactions excited by pump sources with a central wavelength 1401.5 nm, a 10 W peak power, and spectral widths 15 THz and 1 THz (corresponding to temporal widths 0.157 ps and 2.35 ps) in a TFLN waveguide with a length 1.5 mm, an etch depth $d=433$ nm, and a top width $w=3500$ nm are displayed in Figs. \ref{fig:3}(a,b), respectively. The phase matching contour of this nonlinear process is highlighted in Fig. \ref{fig:2}(b). The spectral bandwidth of the JSI using the 15 THz pump source is approximately 210 THz (from 940 nm to 2730 nm), more than an octave frequency range, and with a Schmidt number $K=60$. Using the 1 THz pump yields the same broad spectral bandwidth, however, with a significant higher Schmidt number $K=910$, indicating a much higher-dimensional frequency entanglement.  The estimated photon pair generation rate is $3 \times 10^5~\text{s}^{-1}$ for the latter pump source assuming a 80~MHz repetition rate. Similar SFWM bandwidth (200 THz) with a  Schmidt number $K=860$ can also be obtained using the waveguide highlighted in Fig. \ref{fig:2}(a) with $d=300$ nm and $w=1700$ nm.


The dependence of the Schmidt number on the sampling frequency of the signal and idler photons for different pump bandwidths using the waveguide with $d = 433$ nm and $ w = 3500$ nm is depicted in Figs. \ref{fig:3}(c,d). Clearly, the Schmidt number scales inversely with pump bandwidth; as the pump bandwidth decreases, the Schmidt number increases. The estimated Schmidt numbers using the approximated equation \cite{Zhang07} are portrayed as horizontal dashed lines, showing an excellent agreement with the converged numerical solutions obtained using the SVD technique. A 0.5 THz pump source would require more than 5500 frequency sampling points to reach convergence, which is very computationally intensive. On the other hand, the analytical solution estimate provides a very fast and accurate prediction of the Schmidt number, remarkably useful in the quasi-monochromatic limit. Hence, an exceptionally large  $K \sim 10^8$ is anticipated using a quasi-monochromatic MHz pump source, demonstrating a very high dimensional frequency entanglement achievable via our system. 

\subsection{AlGaAs buried waveguides}
\begin{figure}[t]
\centering\includegraphics[width=12 cm, height=12 cm]{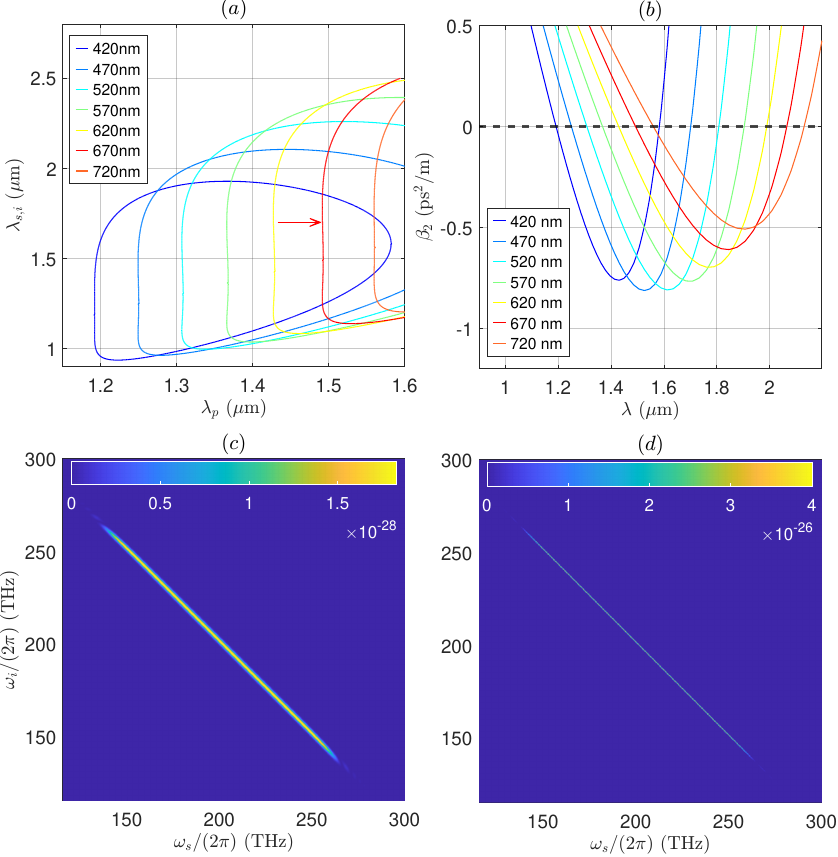}
\caption{(a,b) Phase-matching contours $(\Delta k=0)$ and GVD for SFWM processes in Al$_{0.3}$Ga$_{0.7}$As waveguides with height 308~nm, and varying widths (420--720 nm). (c,d) JSI for SFWM processes triggered by a pump source with a central wavelength 1493.5 nm and spectral widths of 15 THz and 1 THz, respectively, in Al$_{0.3}$Ga$_{0.7}$As waveguide with height 308 nm, width 671 nm, and length 1 mm long. Colorbars represent the probability density in units of Hz$^{-2}$.}
\label{fig:4}
\end{figure}

In this subsection, we examine Al$_{0.3}$Ga$_{0.7}$As waveguides buried in silica cladding, as illustrated in Fig.\ref{fig:1}(b), for spectral entanglement. The phase-matching contours of SFWM processes in different AlGaAs rectangular waveguides with a core height  308 nm and multiple core widths are portrayed in Fig. \ref{fig:4}(a).  All the interacting photons are assumed to be in the waveguide fundamental TE mode. Similar to TFLN waveguides as the waveguide width increases, the spectral flatness of the phase-matching contours broadens and is shifted to longer pump wavelengths. Panel (b) displays the waveguides GVD characterised by small values with 2 ZDWs, where the shorter wavelength determines the optimal pump wavelengths for achieving the broadest SFWM bandwidths, similar to TFLN waveguides.

The output JSI of SFWM processes in an Al$_{0.3}$Ga$_{0.7}$As waveguide with height $h=308$~nm, width $w=670$~nm, and $L=1$ mm is shown in Figs. \ref{fig:4}(c,d), using a 0.5 W pump source at a central wavelength 1493.5 nm and spectral widths of 15 THz and 1 THz, respectively. In both cases, the JSI spans approximately 120~THz, covering the range 1150~nm to 2140~nm, with Schmidt numbers $K=34$ for the 15~THz pump, and $K=516$ for the 1~THz pump. Although these Schmidt numbers are lower in comparison to those obtained in the TFLN waveguides, they still represent exceptionally broad high-dimensional spectral entanglement. The photon-pair generation rate is $3.5 \times 10^6~\text{s}^{-1}$ using the 1~THz pump source  with an order less peak power than in TFLN waveguides because of the higher AlGaAs nonlinear coefficient.


\begin{figure}[t]
\centering
\includegraphics[]{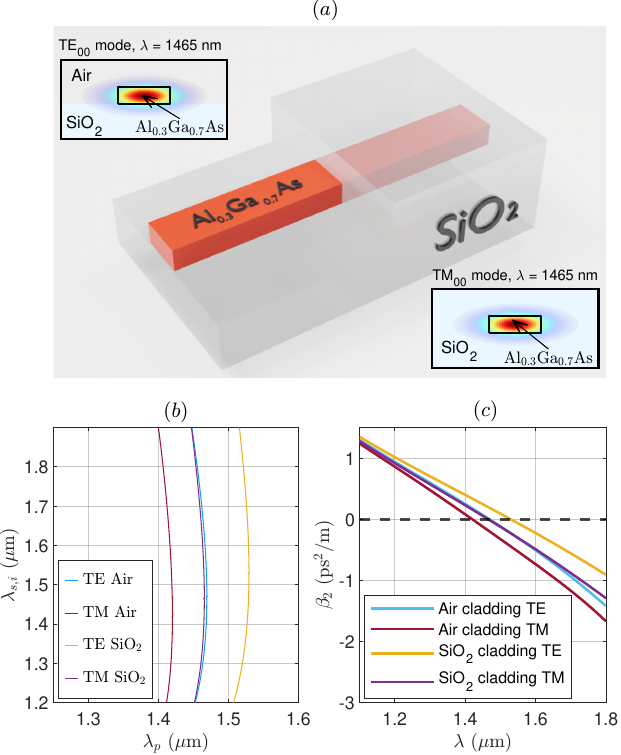}
\caption{(a) Sketch of the hybrid top-cladding Al$_{0.3}$Ga$_{0.7}$As waveguide. (b) Phase-matching curves of SFWM processes with all photons in either TE or TM modes in air- and silica-clad Al$_{0.3}$Ga$_{0.7}$As waveguides with a height 655 nm and  width 755 nm. (c) Dispersion curves of the fundamental TE and TM modes in these waveguides. Insets in (a) illustrate the mode profiles of the fundamental TE mode in air-clad and TM mode in silica-clad waveguides at $\lambda = 1465$ nm wavelength.}
\label{fig:5}
\end{figure}

\section{Broadband polarisation entanglement using nano-photonic waveguides}
In this section, we extend our investigation to study polarisation entanglement using dispersion-engineered AlGaAs waveguides. The aim is the generation of a polarisation-entangled symmetric Bell state, with all the interacting photons in either TE or TM mode, Eq. (\ref{eqn:Bell}). Our approach employs a single core AlGaAs waveguide but with a hybrid cladding made of 2 successive segments, as depicted in Fig. \ref{fig:5}. The top claddings are air followed by silica to enable double-propagating SFWM processes of all photons in TE and TM polarisations, respectively. The mode profiles of the fundamental TE and TM modes in air- and silica-clad AlGaAs waveguides are also portrayed. TFLN waveguides have not been exploited for this application because the broad phase-matching could not be achieved for both all-TE and TM-polarised SFWM processes at the same pump wavelength. Additionally, the nature of the $\chi^{(3)}$ tensor of LiNbO$_3$ would allow additional SFWM processes with unwanted polarisation combinations, deteriorating the entanglement quality.


 
The pump wavelength and waveguide parameters are optimised to ensure that the SFWM processes in the 2 segments of the hybrid waveguide are nearly identical, enabling high concurrence. The phase-matching curves ($\Delta k = 0$) of all TE- and TM-SFWM interactions in air- and silica-clad Al$_{0.3}$Ga$_{0.7}$As waveguides with a height 655 nm and  width 755 nm are displayed in Fig. \ref{fig:5}(b). As shown, the phase matching curves of all-TE and all-TM interactions coincide over a very broad signal/idler bandwidth around pump wavelength $\lambda_p = 1465$ nm in the waveguides with air and silica cladding, respectively. Likewise as portrayed in panel (c), the fundamental TE- and TM-mode dispersion profiles $\beta_2$ in these two waveguide configurations overlap, indicating similar mode profiles, 
group velocities and ZDWs. The mode overlap between the fundamental TE and TM modes in the air and silica claddings exceeds 99.7\% across the entire signal and idler wavelength range, allowing efficient and smooth transition of the modes at the interface between the two sections.

\begin{figure}[t] 
    \centering 
    \includegraphics[scale=0.9]{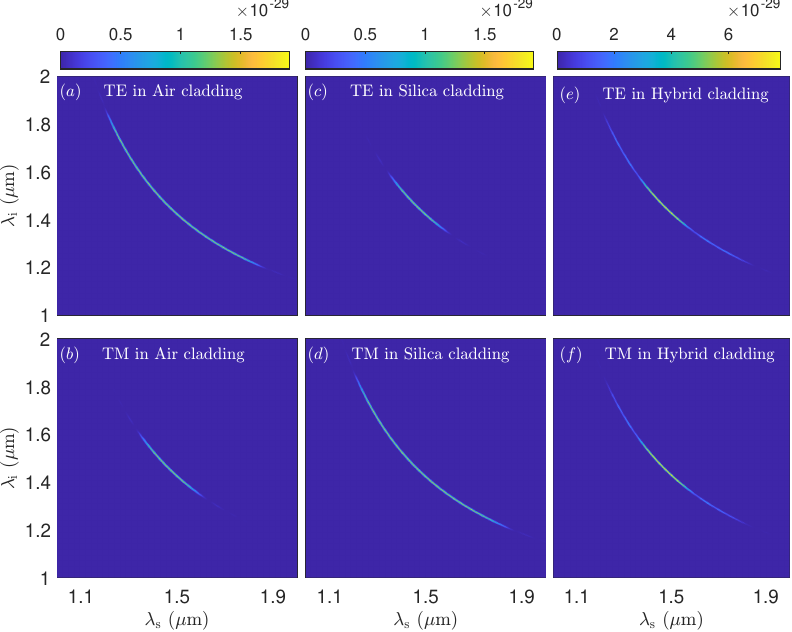} 
    \caption{Joint spectral intensities for SFWM processes in waveguides with air-clad (a,b) silica-clad (c,d), and  hybrid-clad (e,f) using a Gaussian pump source with a peak power of 100 mW, a spectral width of 4 THz (0.588 ps), and a central wavelength 1463 nm. The top and bottom rows are for all-TE and all-TM nonlinear interactions, respectively. The waveguide length is 2 mm long for air- and silica-clad cases and 4 mm long for the hybrid-clad waveguide. The colorbars denote the probability density in units Hz$^{-2}$.}
    \label{fig:6} 
\end{figure}

\begin{figure}[t]
\centering
\includegraphics[width=12 cm,height=12 cm]{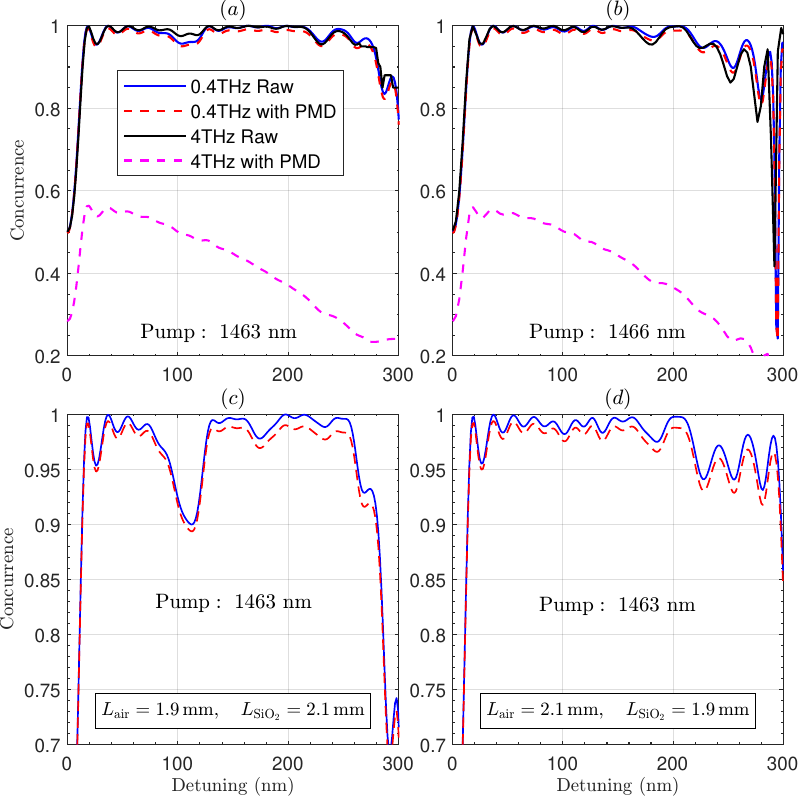}
\caption{Dependence of the concurrence $C$ on the wavelength-detuning from the pump central wavelength. The solid and dashed lines for the raw and PMD-included concurrence, respectively. (a) For pump central wavelength 1463 nm, with two different pump bandwidths 0.4 and 4 THz, and 2 mm long for the air and silica sections. (b) The same as (a) except the pump central wavelength is 1466 nm. (c,d) For a pump central wavelength 1463 nm, pump bandwidth 0.4 THz, and unequal lengths for the air and silica sections, with total waveguide length of 4 mm long.}
\label{fig:7}
\end{figure}

Launching a pump source at a wavelength very close to the ZDW and $45^{\circ}$ polarization angle in the hybrid waveguide would dominantly generate TE- and TM-polarised pairs in the first and second segments of the waveguide, respectively, over a broad range of frequencies. The JSIs for all-TE and all-TM SFWM interactions pumped by a Gaussian pulse with 4 THz (0.588 ps) bandwidth and 100~mW peak power in the air-clad, silica-clad, and hybrid-clad waveguides are depicted in Fig. \ref{fig:6}. The hybrid-clad waveguide is 4 mm long made of 2 mm air-clad followed by 2 mm silica-clad. The generated photon pairs are nearly indistinguishable in frequency and time-arrival at the detectors, since the two SFWM processes exhibit nearly identical joint spectral properties. However, they possess orthogonal polarisations, creating a broadband polarisation entanglement spanning 1175-1750 nm. Although the all-TE and -TM JSIs at the end of the hybrid-clad waveguide are not flat over the entire wavelength range, they have similar spectral distribution, which will lead to high concurrence values.  The non-flatness is due to the air-cladd section allows phase matching of all-TM SFWM interactions over approximately 240 nm around the pump wavelength, whereas the silica-clad section has a similar effect on the all-TE SFWM interactions.  Therefore, these spectral bands effectively experience more efficient photon generation rates over other wavelengths. The average photon-pair generation rate is estimated to be  $ \approx 3.5 \times 10^5~\text{s}^{-1}$.





The dependence of the concurrence that measures the quality of the polarisation entanglement on the wavelength-detuning from the pump central wavelength is depicted in Fig. \ref{fig:7} for  a pump spectral bandwidths 0.4 and 4 THz. Panels (a,b) display the raw concurrence (solid lines) and concurrence including the PMD effects (dashed lines) for two different pump central wavelengths. As demonstrated, the concurrence is very high ($>0.9$) for the 0.4 THz case across a wide detuning range, with negligible PMD effects since $\tau_s \approx \tau_i$ in the proposed structure. As the pump bandwidth widens to 4 THz, the PMD effect significantly degrades the concurrence and rapidly reduces it with increasing the wavelength-detuning.  However, the output spectra of the generated photon pairs are the same for both cases. This allows for mitigating the PMD effects without compromising  the performance of the proposed device via adjusting the pump bandwidth. The concurrence is inevitably degraded within 27 nm around the pump wavelength due to the unwanted contributions of SFWMs with signal and idler in opposite polarisations. This mainly affects a small portion of the S- and E- telecommunication bands. The differential group delay $\tau$ used in calculating the PMD effects is based on the difference between TE-mode in the silica cladding and the TM-mode in the air cladding, which corresponds to the largest possible delay. Panels (c,d) optimise the relative lengths of the air- and silica-clad sections to maximise the concurrence for the 0.4 THz pump source. Using 2.1 mm long air-clad and 1.9 mm long silica-clad would yield >0.93 concurrence across an exceptionally wide detuning range from 14 nm to 290 nm.



\section{Conclusion}
 In this paper, we have shown how integrated nanophotonic waveguides can be dispersion-engineered to achieve ultra-broad frequency and polarisation entanglement, covering multiple telecom bands. Using thin-film lithium niobate (TFLN) and AlGaAs waveguides,  frequency-entangled photon pairs spanning over 940--2730 nm (210 THz) and 1150--2140 nm (120 THz) can be generated, respectively. The characteristic Schmidt number of the entangled source can approach $10^{8}$ for a quasi-monochromatic pump, allowing high-dimensional frequency entangled states.   Although, accessing the full dimensionality of such states requires measurement capabilities beyond current technology, our approach still remarkable since practical high accessible states can still be obtained via controlling the pump bandwidth. Furthermore, using AlGaAs waveguides with hybrid cladding can enable polarisation-entangled photon pairs with high concurrence $>0.93$ across the spectral range 1175--1750 nm. Table~\ref{tab:comparison} benchmarks the performance of our proposed structures against other exisiting platforms in terms of spectral bandwidth and entanglement quality, showing the exceptional performance of our devices. 

\begin{table}[t]
\centering
\caption{Benchmarking Photon-Pair Sources for Broadband Entanglement Generation}
\label{tab:comparison}
\footnotesize  
\setlength{\tabcolsep}{3.5pt} 
\begin{tabular}{|l|l|l|l|c|}
\hline
\textbf{Platform} & \textbf{Bandwidth} & \textbf{Entanglement} & \textbf{Quality Metric} & \textbf{Ref.} \\
\hline
PPLN waveguide & 100 THz (1.2--2 $\mathrm{\mu m}$) & Spectral & Visibility: 98\% & \cite{javid_ultrabroadband_2021} \\
\hline
SiN & 0.7--1.7 $\mathrm{\mu m}$ & Spectral & Not specified & \cite{vijay_sin_2023} \\
\hline
Silica (PCF) & $\sim$1000 nm & Spectral & $K >1.7\times 10^3$ & \cite{garay-palmett_ultrabroadband_2008} \\
\hline
PPLN Superlattice & 90 THz (1.25--2 $\mathrm{\mu m}$) & Spectral & $K \sim 10^7$ & \cite{Zhang07} \\
\hline
TFLN & 210 THz (0.94--2.73 $\mathrm{\mu m}$) & Spectral & $K \sim 10^8$ & This work \\
\hline
Si & 150 nm (1475--1625 nm) & Polarisation & $C\sim$ 0.98 & \cite{sharma_silicon_2022} \\
\hline
Fiber (PPSF) & 24 THz & Polarisation & $C\sim$ 0.91 & \cite{chen_broadband_2021} \\
\hline
AlGaAs & 60 nm  & Polarisation & $C\sim$ 0.91 & \cite{Kultavewuti17} \\
\hline
AlGaAs & 575 nm (1175--1750 nm) & Polarisation & $C>0.93$ & This work \\
\hline
\end{tabular}
\end{table}

 These proposed broadband entangled-photon sources will have significant impact in quantum  applications, including multi-wavelength entanglement distribution, quantum key distribution, and broadband quantum metrology. Their wide spectral coverage will ensure compatibility with existing telecom infrastructure. Whereas their scalability and robustness, for instance against polarisation mode dispersion, will enhance their practical feasibility. Finally, on-chip integration of these sources with pump lasers, and single-photon detectors will ultimately create a pathway to achieve scalable quantum photonic processors.




\begin{backmatter}
\bmsection{Funding}Engineering and Physical Sciences Research Council (EPSRC)
Doctoral Training Partnership (DTP); Institute of Photonics and Quantum
Sciences at Heriot-Watt University.

\bmsection{Disclosures} The authors declare no conflicts of interest.

\bmsection{Data availability} Data underlying the results presented in this paper are not publicly available at this time but may be obtained from the authors upon reasonable request.

\end{backmatter}

\bibliography{sample}

\end{document}